\documentclass[aps,prb,reprint,superscriptaddress]{revtex4-2}
\usepackage{hyperref}
\usepackage{graphicx}
\usepackage{color}
\usepackage{amsfonts}
\usepackage{amsmath}
\usepackage[utf8]{inputenc}
\usepackage[T1]{fontenc}
\usepackage{mathtools}
\usepackage{bbm}
\usepackage[dvipsnames]{xcolor}
\hypersetup{colorlinks=true, urlcolor=blue, citecolor = blue}

\begin{document}

\title{Enhancement of the topological regime in elongated Josephson junctions}

\author{D. Kuiri}
\email{kuiri@agh.edu.pl}
\affiliation{AGH University of Krakow, Academic Centre for Materials and Nanotechnology, al. A. Mickiewicza 30, 30-059 Krakow, Poland.}

\author{P. Wójcik}
\email{pawel.wojcik@fis.agh.edu.pl}
\affiliation{AGH University of Krakow, Faculty of Physics and Applied Computer Science, al. A. Mickiewicza 30, 30-059 Krakow, Poland.}

\author{M. P. Nowak}        
\email{mpnowak@agh.edu.pl}
\affiliation{AGH University of Krakow, Academic Centre for Materials and Nanotechnology, al. A. Mickiewicza 30, 30-059 Krakow, Poland.}

\date{\today}

\begin{abstract}
We theoretically study topological superconductivity in elongated planar Josephson junctions. In the presence of spin-orbit coupling and an in-plane magnetic field, the Josephson junction can enter the topological phase and host zero-energy Majorana bound states over a range of the superconducting phase difference centered around $\pi$, with the span of this range determined by the strength of the magnetic field. We demonstrate that the topological superconducting phase range can be greatly increased by elongation of the junction, which causes an amplification of the Zeeman-induced phase shift of Andreev bound states. We show that the appearance of trivial in-gap states that occurs in elongated junctions can prohibit the creation of Majorana modes, but it can be mitigated by further proximitization of the junction with additional superconducting contacts. The topological transition in this system can be probed by measurements of the critical current and we show that the elongation of the junction leads to a linear decrease of the transition critical magnetic field beneficial for experimental studies.
\end{abstract}

\maketitle

\section{Introduction}
Planar superconductor-normal-superconductor (SNS) junctions have emerged as promising platforms for studying topological superconductivity as they combine necessary spin interactions such as Rashba spin-orbit coupling, a strong Zeeman interaction [provided by the semiconducting normal part typically realized in InAs, InSbAs \cite{PhysRevLett.124.226801, PhysRevLett.130.116203, Fornieri2019, PhysRevLett.130.096202, PhysRevB.107.245304, doi:10.1021/acs.nanolett.1c03520, PhysRevB.93.155402, Moehle2022} two-dimensional electron gases (2DEG)] accompanied by the extended tunability of the system band structure through the variation of the superconducting phase difference $\phi$ between the two superconducting contacts \cite{PhysRevX.7.021032, PhysRevLett.118.107701}. Typically, in hybrid SNS junctions, a superconducting gap with a value similar to that of the parent superconductor is opened by proximitizing 2DEG by a conventional superconductor such as Al \cite{Kjaergaard2016}.

Experiments aimed at the realization of topological superconductivity look for zero-energy Majorana bound states (MBS) that appear at junction edges in the topological phase \cite{PhysRevLett.118.107701}. Typically, tunneling spectroscopy is used to explore the presence of zero-bias conductance peaks \cite{Fornieri2019, PhysRevB.107.245304}, their correlation at the two edges of the system, or a non-local signal characterizing the closing and reopening of the gap \cite{PhysRevB.97.045421, PhysRevB.108.205405}. The system can be tuned in or out from the topological regime mainly by two experimentally accessible knobs: the strength of the Zeeman interaction, which is controlled by the in-plane magnetic field magnitude and by the superconducting phase difference. The latter is induced by embedding the junction in a large superconducting loop, which is threaded by a magnetic flux \cite{PhysRevB.107.245304, Moehle2022}. This method inconveniently suffers from non-zero loop inductance, which translates into phase jumps \cite{PhysRevLett.126.036802,PhysRevB.108.205405} that prevent accessing of the $\phi = \pi$ region, where the MBS form. This in consequence results in the need for MBS to be present in a large phase range, which in turn requires large in-plane magnetic fields. This denies the original feature of the topological SNS junctions, which is the possibility of achieving the topological phase at small magnetic fields \cite{PhysRevX.7.021032}. The strong Zeeman interaction in hybrid SNS systems leads to the appearance of an abundance of trivial in-gap states \cite{PhysRevLett.119.176805, doi:10.1021/acs.nanolett.1c03520}, reduces the induced gap, and the required considerable magnetic field closes the superconducting gap of parent low-critical-field superconductors, such as Al. 

In this paper, we show that the topological regime in the SNS junction can be greatly extended by elongating the junction. However, this comes with a caveat: the formation of transverse modes located in the normal part of the junction, which quickly close the induced gap and destroy the MBS. We propose a way to overcome this problem by further proximitization of the normal region by two superconducting contacts. This kind of configuration has been considered in the context of multiterminal Josephson junctions \cite{PhysRevB.75.174513, PhysRevB.90.075401, PhysRevX.10.031051, PhysRevB.99.075416, Arnault2022, PhysRevB.101.054510, Graziano2022, cayao2024} that have been studied in terms of quasiparticle quartets \cite{Huang2022, PhysRevB.107.L161405, PhysRevB.108.214517}, Cooper pair splitters \cite{PhysRevLett.106.257005}, and superconducting diode effect \cite{Gupta2023, Coraiola2024, correa2024theoryuniversaldiodeeffect}. We show that the topological phase span in the superconducting phase space is linearly dependent on the junction length and that the topological transition can be probed via critical current measurement \cite{PhysRevX.7.021032, PhysRevLett.126.036802, PhysRevB.107.L201405, 10.21468/SciPostPhys.17.4.101, GUARCELLO2024115596}, which also benefits from the enhancement of the topological regime with a linear decrease of the critical magnetic field with junction elongation.

\section{Proof of concept}

\begin{figure}[htp!]
    \centering
    \includegraphics[width = 0.40\textwidth]{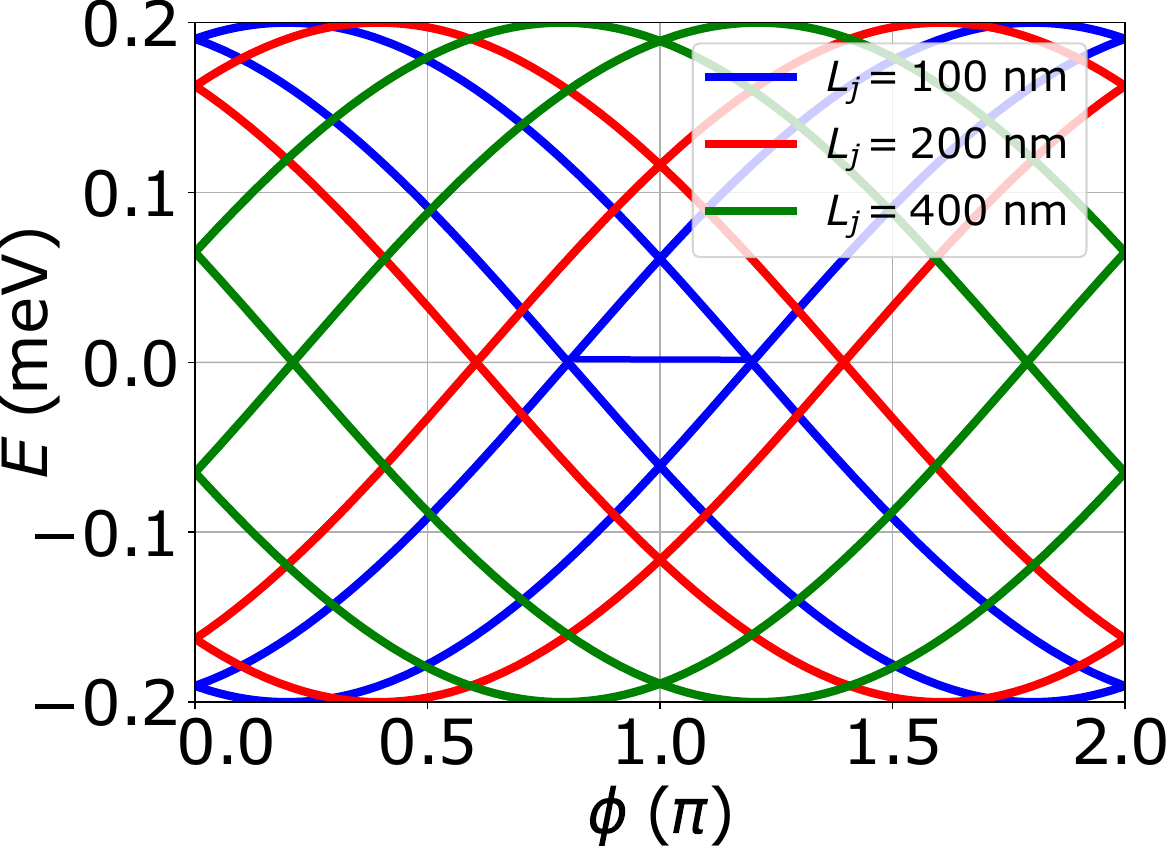}
    \caption{Schematic energy-phase relation of an SNS junction under the Zeeman effect. The blue, red, and green curves show ABS energies for the same magnetic field strength ($B = 0.5$ T) but with an increase in the junction length. The topological regime opens around $\phi = \pi$ with MBS appearing at zero energy (denoted with light lines) and its phase span increases as the junction is elongated.}
    \label{fig:analytical}
\end{figure}

For a perfectly transparent junction the two spin Andreev bound states (ABS) components, dependent on the superconducting gap difference $\phi$, can be described by formula $E_\sigma(\phi) = \pm \Delta\left|\cos[(\phi/2 + \varphi_\sigma)]\right|$, where $\varphi_\sigma = \sigma E_{z}L_j/\hbar v_{F}$, $E_z = g\mu_{B}B/2$ is the Zeeman energy, $\sigma = \pm 1$ corresponds to positive and negative spin components and $v_F = \sqrt{2\mu/m^*}$ is the Fermi velocity \cite{PhysRevLett.67.3836, PhysRevLett.114.227001, PhysRevB.102.165407}. In the calculation of the phase shift we omit the effect of the spin-orbit coupling to the Fermi velocity \cite{PhysRevB.102.165407} as it is small compared to the chemical potential term and has a negligible impact on the obtained results. Here we use $\mu = 5$ meV, $m^* = 0.014m_e$, $\Delta = 0.2$ meV, and $g = 50$ as typical parameters for the InSb SNS junctions \cite{Moehle2022}. As follows from the formula, in the absence of the Zeeman interaction, the ABS energy spectrum is two-fold degenerate. The Zeeman interaction splits the degeneracy and moves apart in phase two branches of ABS states (see the blue curves in Fig.~\ref{fig:analytical}). Between the two branches, around $\phi = \pi$, where the fermion parity is odd, the topological regime emerges, and two MBS are formed at the edges of the junction. Their zero-energy levels are schematically indicated by faint lines in Fig.~\ref{fig:analytical}. 

Crucially, we see that the phase shift of the ABS is proportional to the length of the junction $L_j$---the distance between the superconducting electrodes. In fact, in Fig.~\ref{fig:analytical} we see that the region in which MBS are present is greatly increased when the length of the junction is extended. We can see that for $L_j$ = 400 nm, MBS appear in almost the entire phase region, while for $L_j$ = 100 nm they are present only in the small region around $\phi = \pi$, which can prevent it from being experimentally reached in flux-induced phase-biased Josephson junctions \cite{PhysRevB.108.205405}.

\begin{figure}[ht!]
    \centering
    \includegraphics[width = 0.45\textwidth]{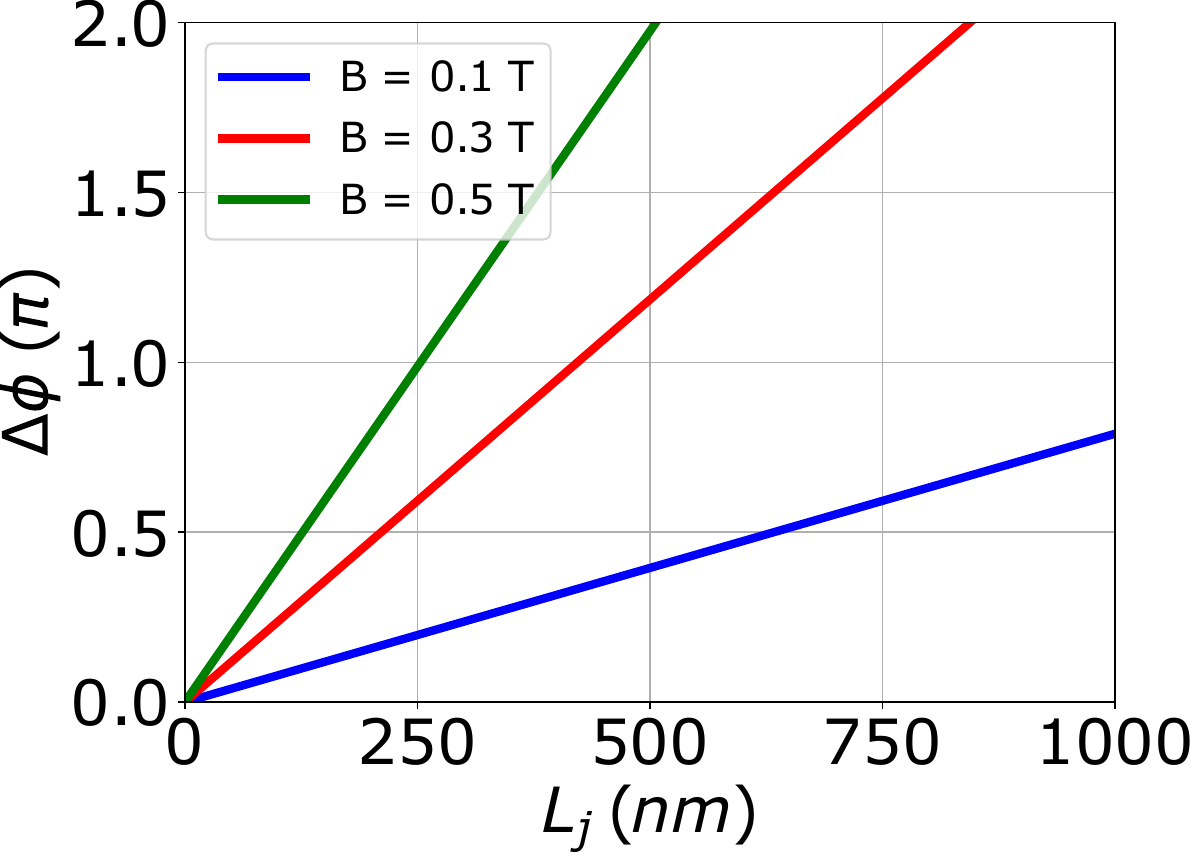}
    \caption{The phase span of the topological regime versus the length of the junction for three values of the magnetic field.}
    \label{fig:analytical_phi_L}
\end{figure}

The size of the odd-parity phase region, which is denoted by the faint lines at zero energy in Fig.~\ref{fig:analytical},  $\Delta \phi = |\phi_{+} -\phi_{-}|$ is determined by the positive and negative roots of $E_\sigma$ = 0 given by $\phi_{\pm} = \pm\chi BL_{j} + \pi(1+2n)$, where $n \in \mathbb{Z}$, resulting in $\Delta \phi=|2\chi BL_j|$ with $\chi = g\mu_{B}/\hbar v_F$. In Fig.~\ref{fig:analytical_phi_L} we show the linear increase in the topological phase span with increasing junction length with the tilt determined by the magnetic field value. It should be noted that the increase of the length of the junction at a given magnetic field can lead to a situation where the topological phase covers the full $2\pi$ phase region with MBS appearing even at $\phi = 0$ in accordance with the phase diagram shown in Ref. \cite{PhysRevX.7.021032}.

In the following sections, we discuss the theoretical framework, numerical simulations, and results that support the proposed idea.

\section{Results}
\subsection{Induced gap---Green's function model}
First, we study the dispersion relation of an SNS junction to understand the effect of the elongation of the junction on the induced gap. We consider a finite system along the superconductor-normal-superconductor direction ($x$) and translation-invariant along the perpendicular direction with a well-defined momentum $k_y$. The SNS junction consists of two S segments corresponding to a proximitized semiconductor flanking a normal (N) semiconducting region \cite{PhysRevB.99.220506}. The system is described by the Hamiltonian,
\begin{equation}
\begin{split}
    H_{0} = &\left( \frac{\hbar^2 k_x^2}{2m^*} + \frac{\hbar^2 k_y^2}{2m^*_\parallel} - \mu \right)\sigma_0 \otimes \tau_z 
    + \frac{1}{2}g\mu_{B}B_y\sigma_y \otimes \tau_0 \\
    &+ \alpha(x)(\sigma_x k_y - \sigma_y k_x)\otimes\tau_z
\end{split}
\end{equation}
written in Nambu basis $\Psi^T = (\psi_{e\uparrow}, \psi_{h\downarrow}, \psi_{e\downarrow}, -\psi_{h\uparrow})$ that represents the electron (e) or hole (h) components with spin up ($\uparrow$) or down ($\downarrow$). $k_{x} = -\iota \partial/\partial x$, $k_y$ is a good quantum number, $\sigma_i$ and $\tau_i$ with $i = x,\;y,\;z$ are the Pauli matrices acting on spin and electron-hole degree of freedom, respectively. The magnetic field and spin-orbit coupling are included only in the normal part of the junction. 

The excitations of the system are obtained from the system Green's function \cite{PhysRevB.84.144522, Stanescu_2013, PhysRevB.99.174511}
\begin{equation}
    G^{-1}(w) = w - H_0 - \sum(w),
\end{equation}
where
\begin{equation}{\label{eq:omega}}
\sum(w) = - \gamma\left[\frac{\omega\sigma_0\otimes\sigma_0 + \Delta(x) \tau_{+} +  \Delta^{*}(x) \tau_{-}}{\sqrt{\Delta(x)^2-\omega^2}}\right], 
\end{equation}
is a self-energy resulting from the uniform coupling of the system to the superconductor \cite{PhysRevB.84.144522, Stanescu_2013} with the pairing term given by
\begin{equation}
\Delta(x) = 
\begin{cases}
    \Delta_{0} & \text{if } x < -L_j/2\\
    0 & \text{if } |x| \le L_j/2\\
    \Delta_{0}e^{i\phi} & \text{if } x > L_j/2\\
\end{cases},
\end{equation}
which corresponds to a two-dimensional electron gas proximitized by two superconducting segments placed on top of the heterostructure \cite{PhysRevLett.124.226801, PhysRevLett.130.116203, Fornieri2019, PhysRevLett.130.096202, PhysRevB.107.245304, doi:10.1021/acs.nanolett.1c03520, PhysRevB.93.155402}. Equation (\ref{eq:omega}) is applicable to energies within the superconducting gap, $|\omega| < \Delta_0$. The density of states is calculated based on the formula
$\rho(\omega) = -\frac{1}{\pi} Im[G(w)]$ \cite{PhysRevB.90.085302}, where we adopt the parameters that correspond to the InSb semiconductor and the Al superconductor with $m^* = 0.014m_e$, $\mu = 5$~meV, $\Delta_0 = 0.2$~meV, $\alpha = 50$~meVnm and $\gamma =0.7$~meV that controls the coupling between the superconductor and the parent semiconductor.

We discretize the Hamiltonian on a square lattice with a lattice spacing of $a = 10$ nm and replace superconducting (proximitized) contacts with finite segments with lengths that exceed the coherence length (estimated to be $\xi \simeq 1\;\mu$m), with $L_{SC}=2\;\mu$m. Since in our calculation we use a uniform chemical potential, for a proper description of Andreev scattering at the NS interface \cite{PhysRevB.59.10176}, we introduce an anisotropic mass in the proximitized contacts with the effective mass in the direction parallel to the interface $m^*_\parallel = 1000 m^*$ \cite{short_junctions}. The results were obtained in part with the use of the Kwant package \cite{kwant} and the numerical code behind the calculations is available in an online repository \cite{code}.

\begin{figure}[ht!]
    \centering
    \includegraphics[width = 0.45\textwidth]{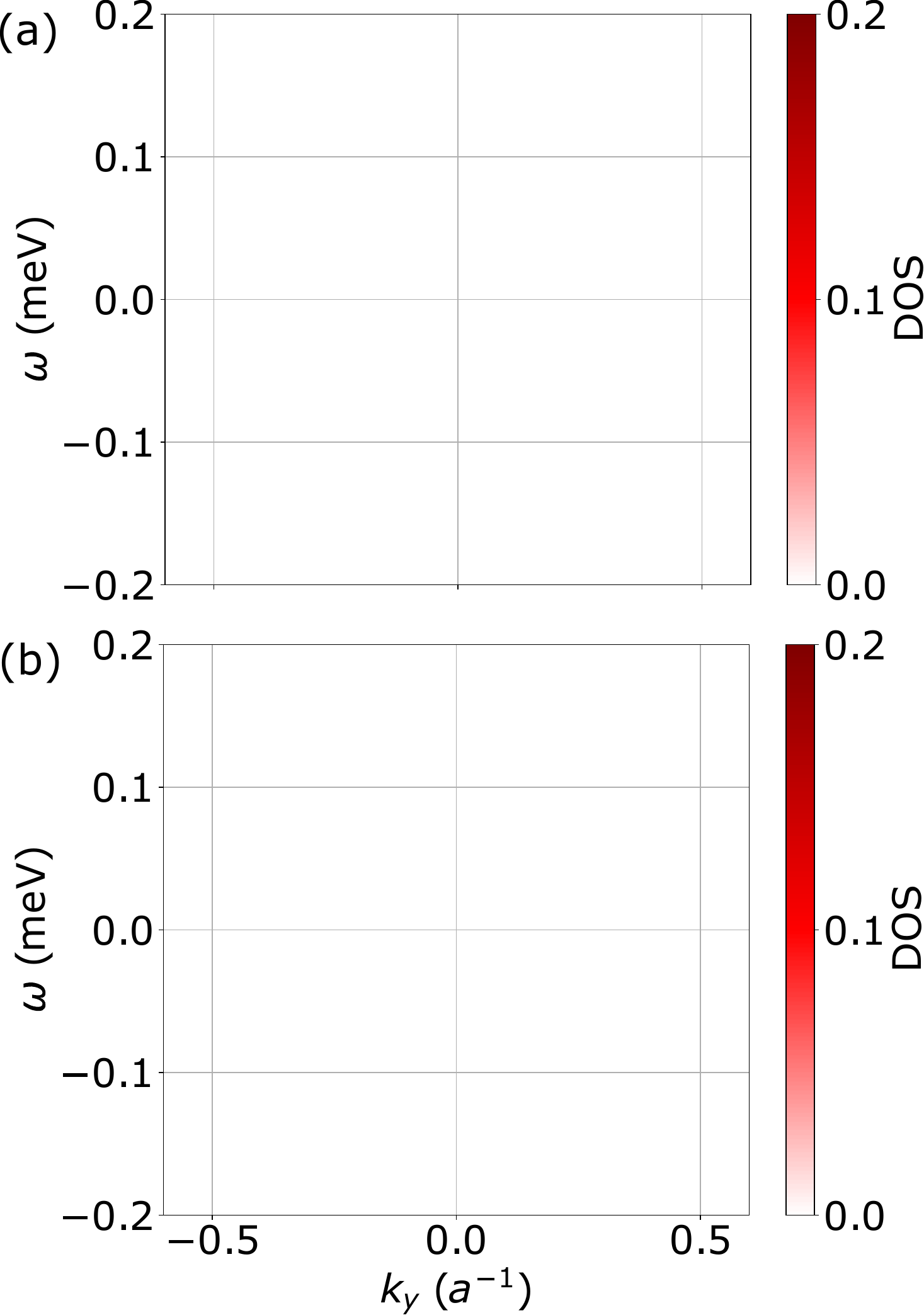}
    \caption{Density of states of a junction invariant along the NS interface  as a function of transverse momentum $k_y$ for the junction length (a) $L_j =100$ nm and (b) $L_j=500$ nm for $\phi=0$ and $B = 0$.}
    \label{fig:in_gap_states}
\end{figure}

In Fig. \ref{fig:in_gap_states} we show the calculated densities of states versus $\omega$ and $k_y$ that visualize the junction spectrum for $L_j = 100$ nm (a) and $L_j = 500$ nm (b) at zero magnetic field and $\phi=0$. We observe the appearance of an induced gap with magnitude close to 200~$\mu$eV and an abundance of quasiparticle bands outside the gap that originate from the wide superconductors. In contrast to the $L_j = 100$ nm results [Fig. \ref{fig:in_gap_states} (a)], where the gap is opened in the whole Brillouin zone, for $L_j = 500$ nm [Fig. \ref{fig:in_gap_states} (b)] there are a great number of states with large momenta $k_y$ localized within the induced gap. As we shall see, those states that appear upon elongation of the junction constitute an obstacle in enhancing the topological gap by elongation of the junction in the simple two-terminal SNS geometry.

\subsection{Integrated superconductivity}
For further analysis of the in-gap states and to numerically approach finite junctions with realistic sizes, determined by the junction length $L_j$ and width $W_j$ (the size of the junction along the superconducting contacts), in the next step we move on to a less numerically demanding model, where the induced superconductivity is integrated into the regions surrounding the normal part of the junction \cite{PhysRevB.97.045421, PhysRevB.108.205405}.

The system Hamiltonian becomes
\begin{equation}\label{eq:1}
\begin{split}
       H &= \left( \frac{\hbar^2 k_x^2}{2m^*} + \frac{\hbar^2 k_y^2}{2m^*} - \mu \right)\sigma_0 \otimes \tau_z + \frac{1}{2}g(x)\mu_{B}B\sigma_y \otimes \tau_0 \\
       &\quad + \alpha(x)(\sigma_x k_y - \sigma_y k_x)\otimes\tau_z + \Delta(x) \tau_{+} +  \Delta^{*}(x) \tau_{-}.
\end{split}
\end{equation}
with $\tau_{\pm} = (\tau_{x} \pm \iota \tau_y)/2$ and  $\tau_x = \sigma_0\otimes\sigma_x$, $\tau_y = \sigma_0\otimes\sigma_y$.\\
The Hamiltonian is discretized on a square mesh with lattice constant $a = 10$ nm. All the system parameters remain the same as in the former section.

\begin{figure}[htp!]
    \centering
    \includegraphics[width = 0.5\textwidth]{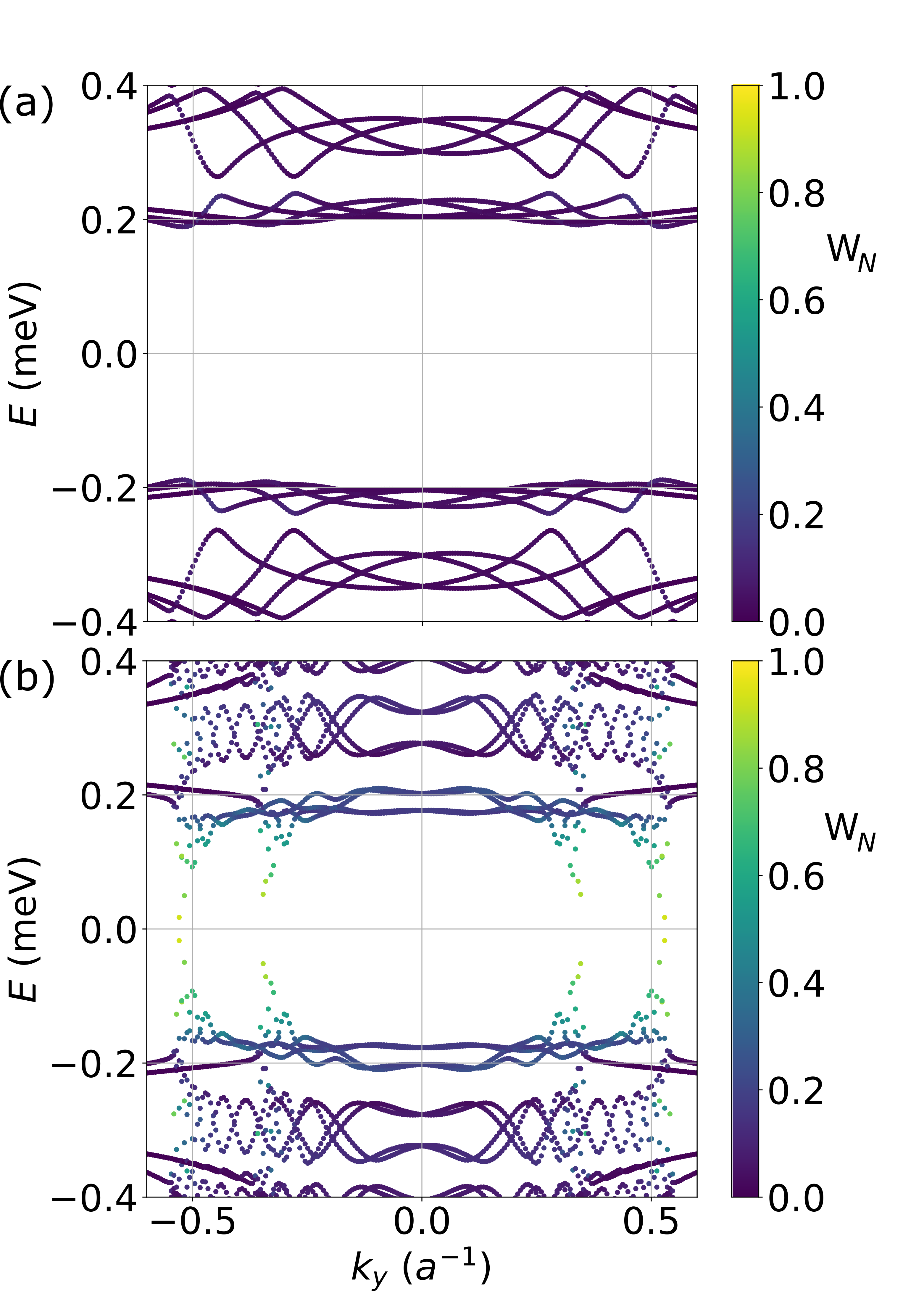}
    \caption{Dispersion relation for translational-invariant SNS junction for (a) $L_j = 100$ nm and the (b) $L_j =500$ nm. The colors denote $W_N$, the weight of the wave function localized in the normal region.}
    \label{fig:dispersion}
\end{figure}

First, we consider a translational invariant system as in the section above. In Fig. \ref{fig:dispersion} we show two dispersion relations for a system with $L_j = 100$ nm (a) and $L_j = 500$ nm (b). As previously, we observe appearance of in-gap states for the elongated junction, which justifies the application of the integrated superconductivity approximation. In Fig. \ref{fig:dispersion} the colors denote the weight of the wave function localized in the normal region of the junction calculated as $W_N = \int_{-L_j/2}^{L_j/2} |\Psi(x, k_y)|^2dx$. For $L_j = 100$ nm, we observe a pronounced superconducting gap open for all $k_y$ values. The bands with the energies outside the induced gap correspond to the quasiparticle states residing mostly in the superconducting leads with small $W_N$ values denoted by a dark blue color. The situation is strikingly different in the elongated junction whose spectrum we plot in Fig. \ref{fig:dispersion} (b). Here, the gap closes for larger $k_y$ values with modes that reside mainly in the normal part of the junction with pronounced $W_N$ values. 

\begin{figure}[ht!]
    \centering
    \includegraphics[width = 0.4\textwidth]{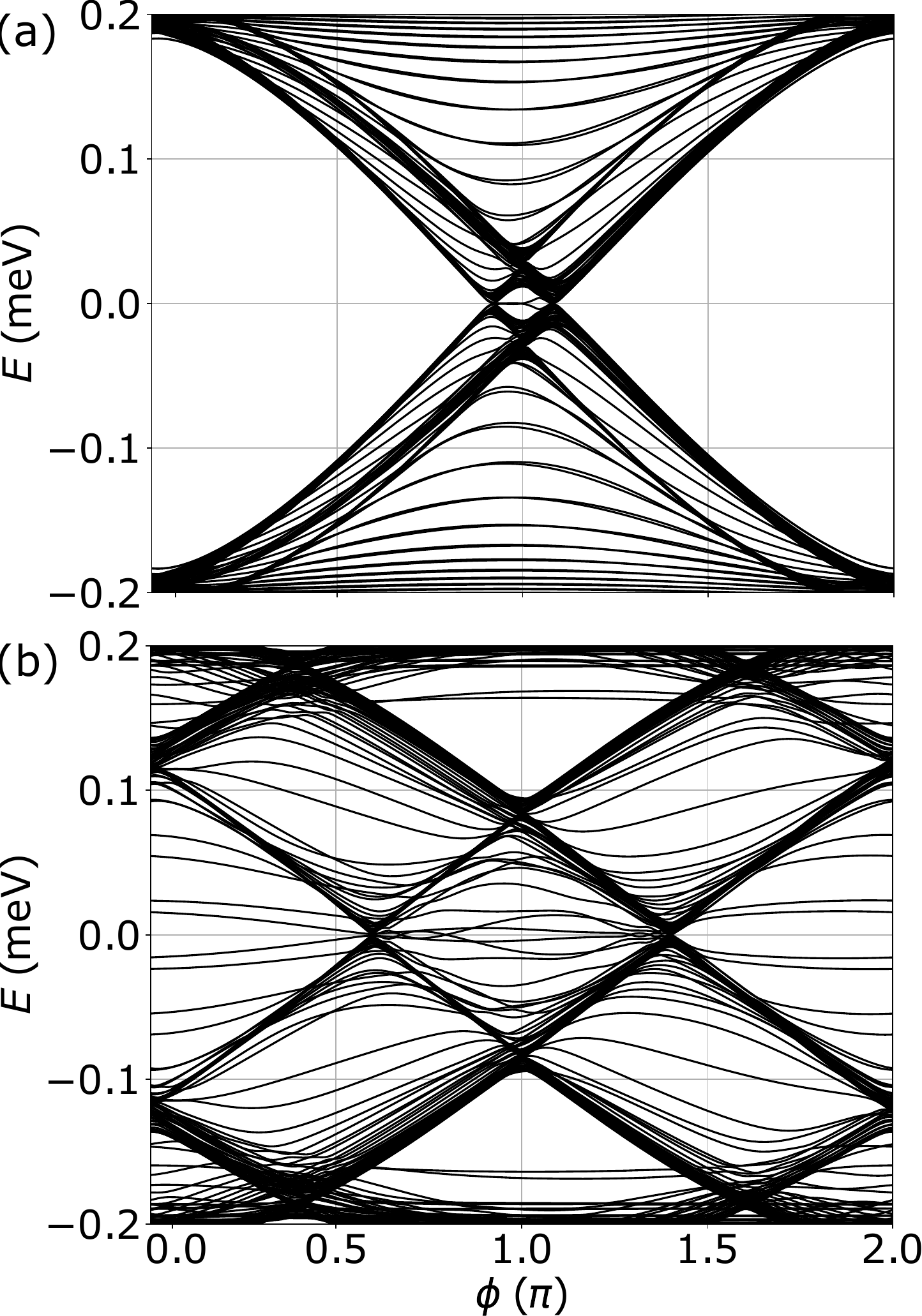}
    \caption{ABS energies versus the superconducting phase difference for a finite-size SNS junction with the length (a) $L_{j} =100$ nm and (b) $L_{j} =500$~nm. The result are obtained for $B=0.2$~T and the width of the junction $W_j = 2000$ nm.}
    \label{fig:ABS}
\end{figure}

The presence of these high $k_y$ in-gap states has immediate consequences for the ABS structure of the finite-size SNS which we show in Fig. \ref{fig:ABS}. Here, the junction length $L_j$ is 100 nm (a) and 500 nm (b) with the width of the systems $W_j = 2000$ nm. In panel (a), we see that the spectrum consists of two spin-split branches of ABS and, between them, a small region with zero-energy states corresponding to MBS in accordance with the proof-of-concept model of Fig. \ref{fig:analytical}. The situation for the elongated junction depicted in panel (b) is starkly different. We again observe two spin-split branches, this time with much larger splitting due to the amplification of the Zeeman effect discussed earlier. However, in the spectrum, there is an abundance of in-gap states whose energies are nearly phase independent. This is in line with the results of Fig. \ref{fig:dispersion}(b) where we saw that the in-gap modes are localized mostly in the normal region and therefore they are weakly affected by change of the phase difference in the nearby superconducting leads. Crucially, those states, mostly localized in the normal regime, couple the two junction edges, making it impossible to form the MBS, which is accompanied by a lack of zero-energy states and a lack of visible induced gap in the topological phase. 

\subsection{Proximitizing the in-gap modes}
Recently, it has been proposed that a zigzag configuration could help mitigate the adverse effects caused by the in-gap modes \cite{PhysRevLett.125.086802}. Here we propose a different strategy which is simpler in terms of the required precision in shaping the superconducting contacts and also avoids the limit of the zigzag junction length set by reopening the straight paths when the separation between the superconductors increases---important in the context of the enhancement of the topological regime.

\begin{figure}[ht!]
    \centering
    \includegraphics[width = 0.45\textwidth]{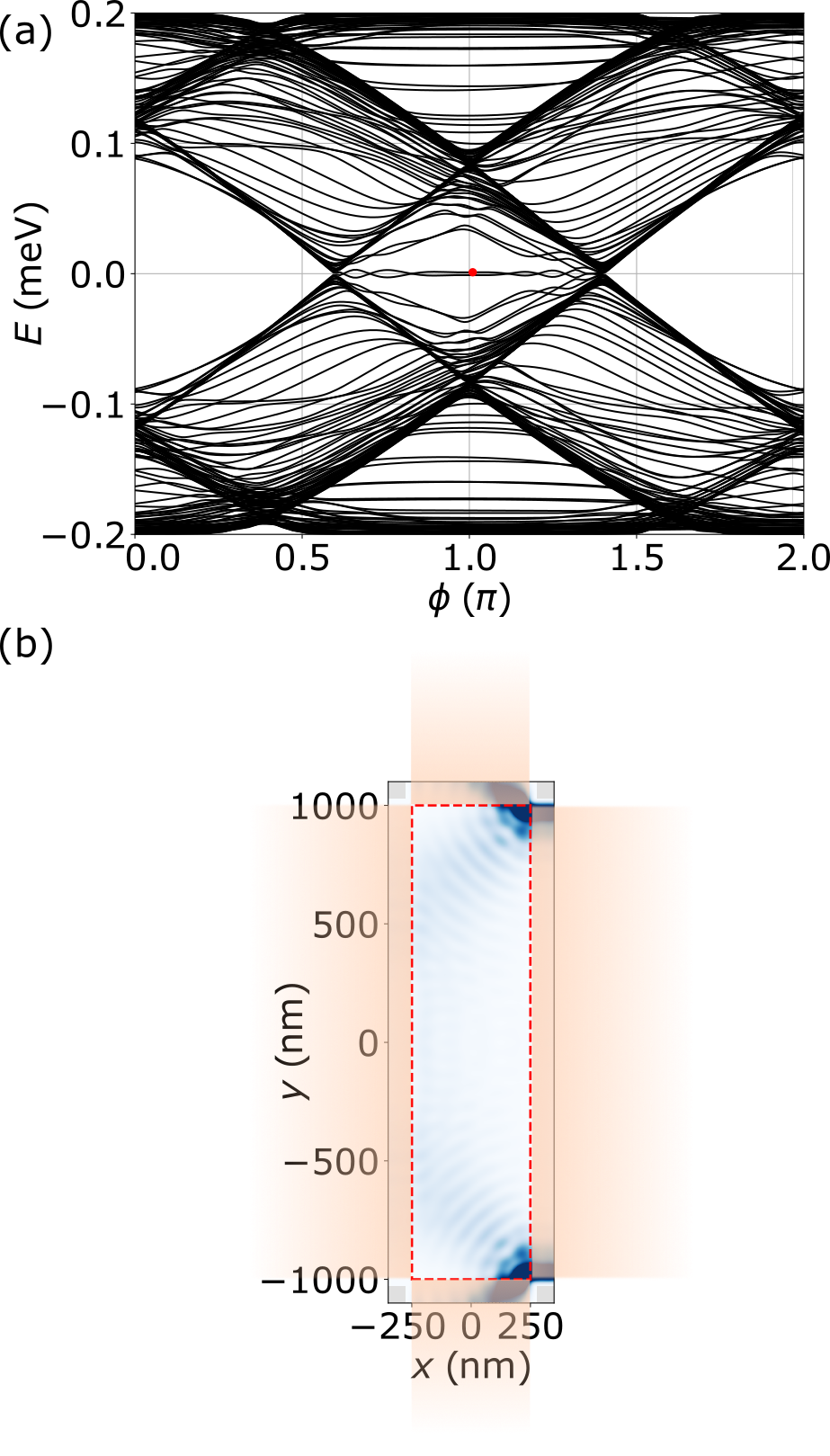} 
    \caption{(a) ABS spectrum for an elongated junction with the length $L_j=500$ nm for $B = 0.2$T with extra two superconducting contacts at the bottom and the top of the junction. (b) Scheme of the extended junction with both left and right and top and bottom superconducting contacts (orange). The blue color map represents the probability density for the MBS denoted with the red dot in (a).}
    \label{fig:gap-reopening}
\end{figure}

Since the in-gap modes have large momenta parallel to the superconducting interfaces and are localized mainly in the normal region, they correspond to quasiparticles traveling between the normal edges of the junction. Hence, we place two additional superconducting contacts at the top and bottom edges of the junction [see Fig. \ref{fig:gap-reopening}(b)] that force Andreev reflections of the vertically propagating quasiparticles, which in turn leads to the reopening of the induced gap. We assume the same superconducting gap value in the top and bottom contacts as in the left and right leads (with the phase $\phi=0$), and the length of 2000 nm that exceeds the value of the superconducting coherence length.

Induced proximitization of the in-gap modes has direct consequences for the ABS spectrum of the junction. In Fig. \ref{fig:gap-reopening}(a) we plot the eigenspectrum of the junction and observe the removal of phase-independent ABS states appearing with energies $|E| < 0.1$~meV in Fig. \ref{fig:ABS}(b). Most importantly, the lack of localized states spreading in the bulk of the normal region allows now for creation of zero-energy modes which correspond to spatially delocalized MBS whose probability density we plot with blue color in Fig. \ref{fig:gap-reopening}(b). It is worth mentioning that the superconducting phase difference, which is implied by setting $\phi \ne 0$ on the right terminal, also breaks the inversion symmetry with respect to $x=0$ and causes a non-symmetric localization of the MBS at the corners of the junction.

\begin{figure}[ht!]
    \centering
    \includegraphics[width = 0.9\linewidth]{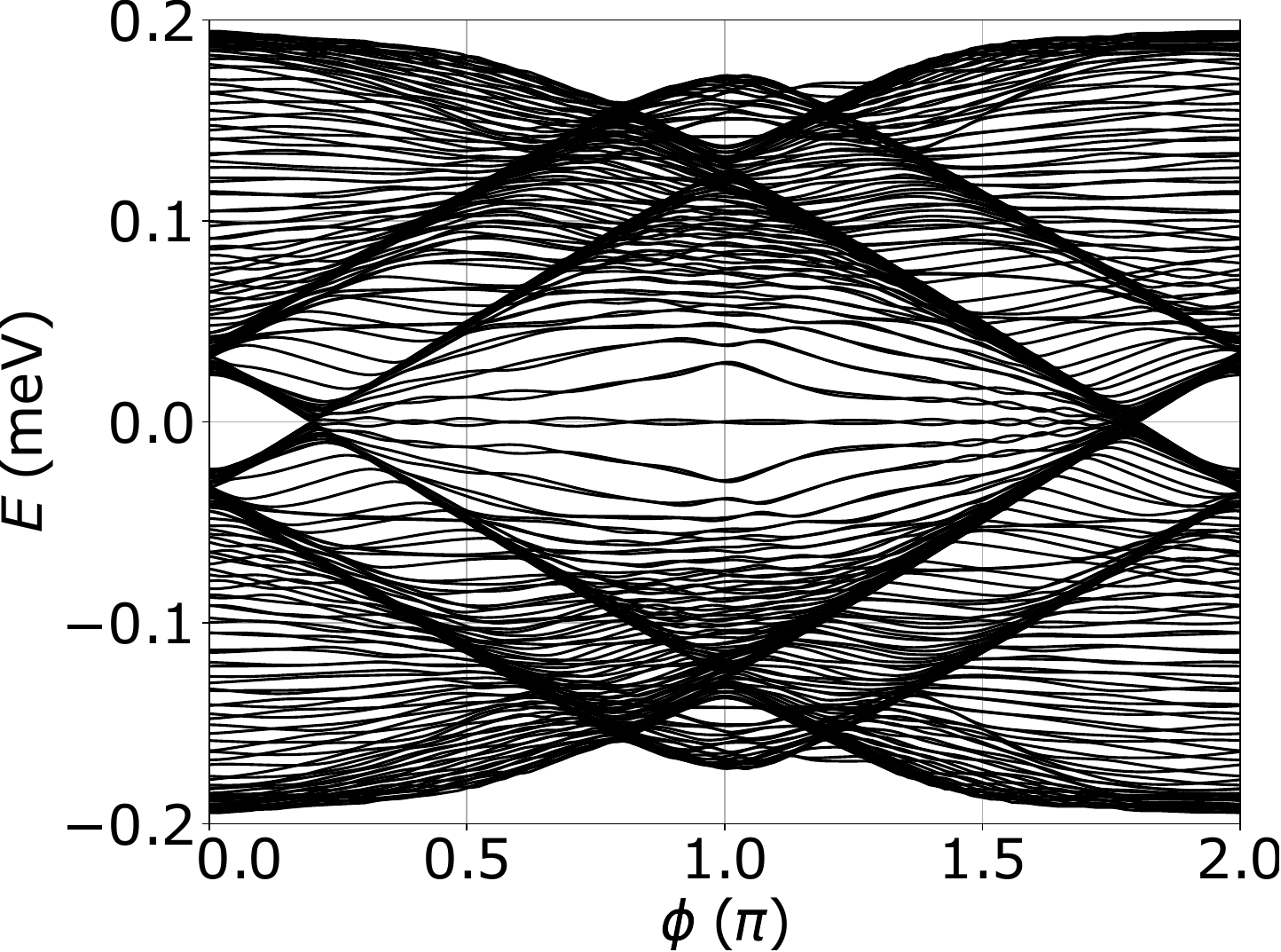} 
    \caption{ABS spectrum for a junction of length $L_j =1000$ nm and $B=0.2$ T.}
    \label{fig:long}
\end{figure}

The elongation of the junction eventually leads to a situation in which the junction length becomes comparable to the superconducting coherence length. In this situation, the ABS structure undergoes a transition, adopting a linear configuration \cite{PhysRevLett.105.077001}. In Fig. \ref{fig:long} we show spectrum of a junction of length $L_j = 1000$ nm $\simeq \xi$. We observe a pronounced topological gap open in the system, and the presence of nearly-degenerate zero energy state almost in the entire phase range despite the small magnetic field value of $B = 0.2$~T, compatible with the prediction of Fig. \ref{fig:analytical}.

\subsection{Possible measurement of topological transition}

The presence of the top and bottom superconducting electrodes prohibits performing local \cite{PhysRevLett.124.226801, Fornieri2019} or non-local \cite{PhysRevB.97.045421, PhysRevB.108.205405} tunneling spectroscopy to probe the ABS structure and the closing and reopening of the gap. However, the topological transition in the ground state of the junction can be investigated by tracing the critical current of the junction $I_c = \mathrm{max}_\phi{I(\phi)}$ (with $I(\phi) = -\frac{e}{\hbar} \sum_{E_{i}>0} {\frac{\partial E_i}{\partial \phi}}$) in an external magnetic field \cite{PhysRevX.7.021032}. $E_i$ are either obtained as eigenvalues of the Hamiltonian Eq. \ref{eq:1} or for the case of analytical approach are calculated by taking $\sum E_i(\phi) = E_+(\phi) + E_-(\phi)$ from the model of Section II with the resulting current scaled to be comparable to the one obtained numerically. The numerical results are obtained for the junction with the top and bottom superconducting contacts attached. 

\begin{figure}[ht!]
    \centering
    \includegraphics[width = 0.45\textwidth]{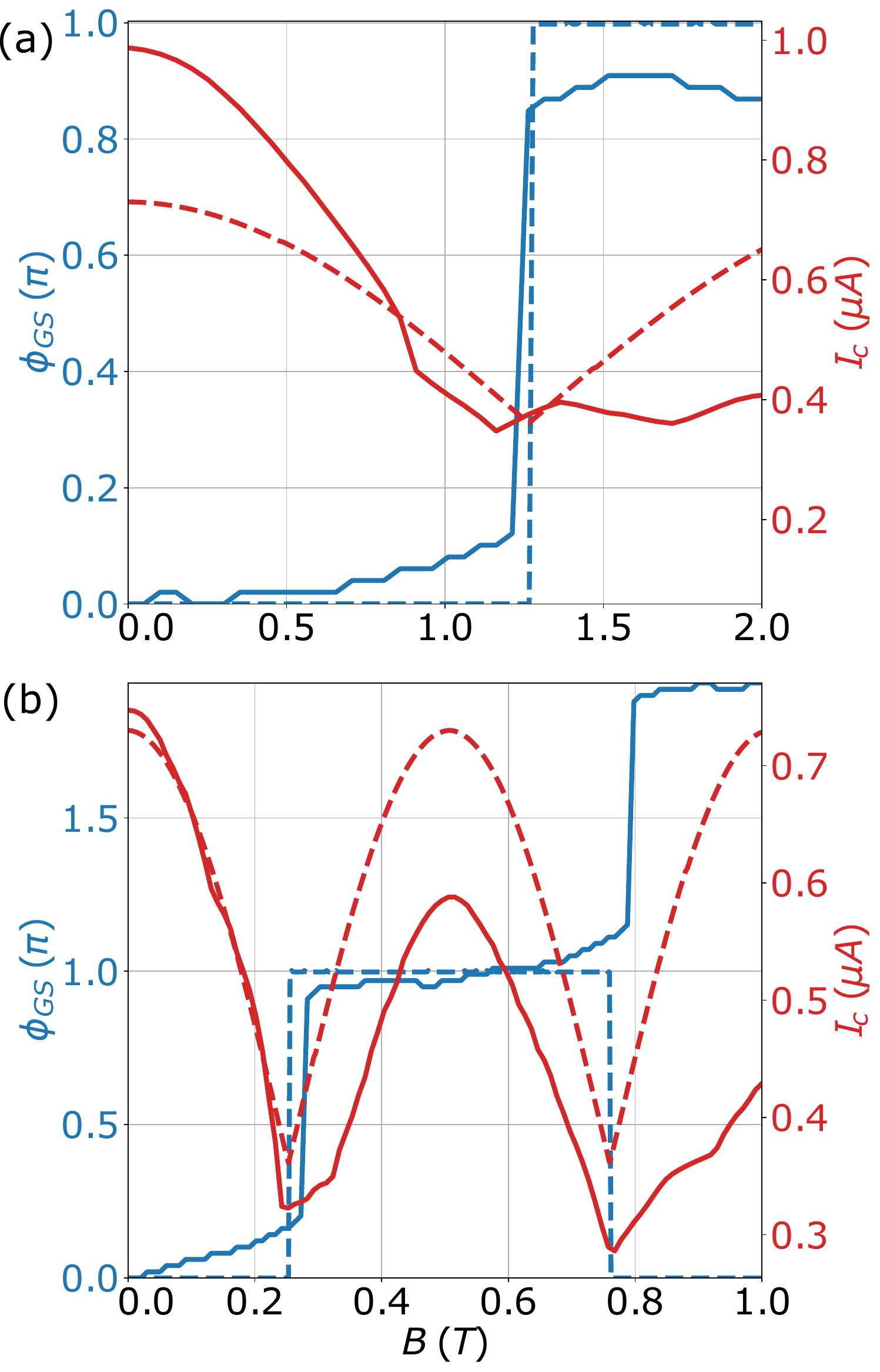} 
    \caption{Superconducting phase difference leading to the ground-state of the junction (blue) and critical current of the junction (red curves), showing minima at phase transitions. The results are obtained for junctions with length (a) $L_j = 100$ nm and (b) $L_j = 500$ nm. Solid lines represent numerical calculations, while dashed lines correspond to analytical results.}
    \label{fig:critical-current}
\end{figure}

In Fig. \ref{fig:critical-current} we show with red solid lines the critical current and with blue solid lines the ground-state phase $\phi_{GS}$ which results in the minimization of the SNS junction energy calculated as the sum over energies of ABS states $E_j(\phi) = \sum_{E_i < 0} E_i(\phi)$. We see that as the critical current reaches the minimum, the ground state of the junction switches from the trivial regime (with a phase difference close to 0) to the topological regime (with the phase difference $\pi$). Crucially, the switching magnetic field is greatly reduced in the elongated junction, as can be seen comparing the panel (a) obtained for $L_j = 100$~nm with the panel (b) calculated for $L_j = 500$~nm. In the same figure with dashed lines we show the ground-state phase and the critical current obtained from the analytical model for ABS spectrum of Fig. \ref{fig:analytical}.

Moreover, from the analytical model for the Zeeman-split ABS we can estimate the magnetic field at which the junction switches its ground-state phase, which occurs at $\sum_\sigma \Delta|\cos(\varphi_\sigma)| = \sum_\sigma \Delta|\cos(\pi/2 + \varphi_\sigma)|$ as
\begin{equation}
    B_c = n \frac{h v_f}{4 L_j g \mu_B},
\end{equation}
with $n \in \mathbb{Z}$ as previously considered for Zeeman-driven Fulde-Ferrell-Larkin-Ovchinnikov-like mechanism that through a spatial variation of the superconducting pairing leads to a minimum in the critical current \cite{Hart2017, PhysRevLett.126.036802}. We see that the linear scaling of the span of the topological region is translated into a linear decrease of the critical field necessary to achieve the topological phase when probing the junction by the critical field.

\section{Discussion }
The typical length of the junctions studied experimentally so far was around 100 nm \cite{PhysRevLett.124.226801, PhysRevLett.130.116203, Fornieri2019, PhysRevLett.130.096202, PhysRevB.107.245304, doi:10.1021/acs.nanolett.1c03520, Moehle2022}, while the measured mean free path was reported to be 600 nm for InAs systems \cite{PhysRevLett.130.116203, Fornieri2019, PhysRevLett.130.096202, PhysRevB.107.245304}, suggesting that the junction length could be increased several times without reducing its transparency significantly and allowing for a several-times increase in the span of the topological region and multiple reductions of the critical magnetic field for the topological transition probed by the critical current. Another possible limit in the achievable junction length is the phase coherence length. However, it is typically much larger than the mean free path as found experimentally for these kinds of heterostructures \cite{PhysRevB.93.155402, PhysRevB.95.155307, Lee2019}, so whenever the junction is clean enough so that the mean free path is longer than the length of the junction, the phase coherence should not be a constraint.

It is also worth comparing the junction length with another length scale present in those systems, namely the spin-orbit coupling length $l_{\mathrm{SO}} = \hbar^2/m^*\alpha$, which for the Rashba-type interaction, to some extent can be controlled by gating the system. For considered here $\alpha = 50$ meVnm $l_{\mathrm{SO}} = 109$ nm and clearly, the junction length exceeds $l_{\mathrm{SO}}$, yet the topological phase is fully developed. We have verified that increasing the spin-orbit coupling strength up to 100 meVnm results in a minimal variation of the energy structure, suggesting that the spin-orbit length does not impose a constraint on $L_j$. It specifically does not affect the topological transition points. This is also expected in analogy to a single superconductor system (such as a nanowire proximitized by a superconducting shell) where the topological transition criterion is $E_z^2 > \Delta^2 + \mu^2$ and does not include the spin-orbit coupling term. It is rather expected that the strength of the spin-orbit interaction affects the spatial extent of the Majorana bound state such that for the systems with stronger spin-orbit coupling the Majorana bound state coherence length decreases \cite{PhysRevB.97.045419}, therefore systems with stronger spin-orbit interactions would be desirable in terms of realization of topological Josephson junctions with limited width (the span in $y$ direction).

\section{Summary and conclusions}
We studied topological superconductivity in elongated planar Josephson junctions. We showed that the increase of the separation between the superconducting contacts causes the amplification of the Zeeman effect affecting the Andreev bound-state spectrum. As the junction becomes longer, at a constant magnetic field, the phase separation between the Andreev bound states increases, extending the phase-span of the topological region. However, as the length increases the junction also becomes populated with trivial in-gap states which prohibit creation of Majorana bound states. We show that those in-gap states can be removed from the spectrum by further proximitizing the junction with two additional superconducting contacts, which restores the topological gap and edge-mode Majorana states. Our numerical study complemented with an analytical model shows that the amplification of the phase-span of the topological regime linearly depends on the junction length. We show that the topological transition can be probed by critical current measurement which also benefits from linear reduction of the critical field, at which the transition occurs as the length of the junction is increased. Furthermore, the presence of Majorana edge modes can be further verified in this type of system by local probing of the density of states \cite{PhysRevB.102.085414} or the spin structure of the edge modes \cite{PhysRevB.102.085411}.

\section*{Acknowledgements}
This work was supported by the National Science Center (NCN) Agreement No. UMO-2020/38/E/ST3/00418. We gratefully acknowledge Polish high-performance computing infrastructure PLGrid (HPC Center: ACK Cyfronet AGH) for providing computer facilities and support within computational Grant No. PLG/2024/017374.

\bibliography{Ref.bib}
\end{document}